\documentclass{aa}
\usepackage{txfonts}
\usepackage{graphicx}
\usepackage{psfig,longtable,lscape}
\usepackage{epsfig}
\usepackage{amsmath}
\usepackage[ ]{natbib} 
\usepackage{multirow}
\bibpunct{(}{)}{;}{a}{}{,} 

\begin{document}

\title{Detailed study of the X-ray and Optical/UV orbital ephemeris of 
X1822--371}
  


\author{R. Iaria\inst{1}, T. Di Salvo\inst{1}, L. Burderi\inst{2}, 
A. D'A\'\i\inst{1}, A. Papitto\inst{2},  A. Riggio\inst{2,4}, 
 N. R. Robba\inst{1}}
 
\offprints{R. Iaria, \email{iaria@fisica.unipa.it}}

\institute{Dipartimento di Fisica,
Universit\`a di Palermo, via Archirafi 36 - 90123 Palermo, Italy
\and 
Dipartimento di Fisica, Universit\`a degli Studi di Cagliari, SP Monserrato-Sestu, KM 0.7, Monserrato, 09042 Italy
  \and
  INAF, Istituto di Astrofisica Spaziale e Fisica cosmica di Palermo, 
via U. La Malfa 153, I-90146 Palermo, Italy
  \and 
 INAF, Osservatorio Astronomico di Cagliari, Poggio dei Pini, Strada 54, 09012 Capoterra (CA), Italy
} 

\date{Received 25 May 2011/ Accepted 25 July 2011}

\abstract
{}
{Recent studies of the optical/UV and X-ray ephemerides of X1822-371
  have found some discrepancies in the value of the orbital period
  derivative.  Because of the importance of this value in
  constraining the system evolution, we comprehensively analyse all
  the available optical/UV/X eclipse times of this source  to
  investigate the origin of these discrepancies.}
{We collected all previously published X-ray eclipse  times from 
  1977 to 2008, to which we added the eclipse  time observed by Suzaku
  in 2006. This point is very important to cover the time gap between the last
  RXTE eclipse time (taken in 2003) and the most recent Chandra 
  eclipse  time (taken in 2008). 
  Similarly we collected
  the optical/UV eclipse arrival times covering the period from 1979
  to 2006, adding a further eclipse  time taken on 1978 and
  updating  previous  optical/UV ephemeris.  We compared the
  X-ray  and the optical/UV ephemeris, and finally derived a new ephemeris 
of the
  source by combining the eclipse arrival times in the X-ray and
  optical/UV bands.  }
{The X-ray eclipse time delays calculated with respect to a constant
  orbital period model display a clear parabolic trend, confirming that
  the orbital period of this source constantly increases at a rate of
  $\dot{P}_{\rm{orb}} =1.51(7) \times 10^{-10}$ s/s.  Combining the
  X-ray and the optical/UV data sets, we find that $\dot{P}_{\rm{orb}}
  =1.59(9) \times 10^{-10}$ s/s, which is compatible with the X-ray orbital
  solution.  We  also investigate the possible presence of a
  delay of the optical/UV eclipse with respect to the X-ray eclipse,
  finding that this delay may not be constant in time.  In particular,
  this variation is compatible with a sinusoidal modulation of
    the optical/UV eclipse arrival times with respect to the long-term
    parabolic trend. In this case, the optical/UV eclipse should lag
    the X-ray eclipse and the time-lag oscillate about an
    average value.}
{ We confirm that the  orbital period derivative is three orders of
  magnitude larger than expected from conservative mass
  transfer driven by magnetic braking and gravitational radiation.   
  }

\titlerunning {Detailed study of  the orbital ephemeris of 
X1822--371}
\authorrunning {R.\ Iaria et al.}

\maketitle

\section{Introduction}

X1822-371 is an eclipsing compact binary system with a period of 5.57
hr hosting a 0.59 s X-ray pulsar.  Several authors have reported
new orbital ephemeris of the source using observations performed in
different energy bands.  \citet[hereafter BU10]{Burderi_2010} analysed
X-ray data of X1822-371 covering the period from 1996 to 2008  to
determine the eclipse times of the source and improved the previous
X-ray ephemeris of X1822-371 reported by \citet[hereafter
PA00]{parmar2000} that covered the period from 1977 to 1996. BU10
added their data to those used by PA00 finding a positive derivative
of the orbital period of $(1.499 \pm 0.071) \times 10^{-10}$ s/s that
is compatible with the previous one given by PA00 but with a smaller
associated error.

\citet[hereafter BA10]{Bayless2010} obtained the optical/UV ephemeris
of X1822-371 using data covering the period from 1979 to 2006. They
obtained a value of the orbital period derivative of $(2.12 \pm 0.18)
\times 10^{-10}$ s/s, which  is compatible with that reported by PA00
 but slightly larger than the value proposed by BU10.

 \citet[hereafter JI11]{Ji2011}, using the X-ray eclipse
 arrival times reported by PA00 and the eclipse arrival times inferred
 by the two Chandra/HETG observations of X1822-371 performed in 2000
 (Obs ID: 671) and in 2008 (Obs ID: 9076 and 9858), already included in
 the work of BU10, estimated a value of the orbital period derivative
 of $(0.83 \pm 0.16) \times 10^{-10}$ s/s, with the error at the  90\%
 confidence level, almost a factor of two smaller than the value reported
 by BU10.

 We summarise the values of the eclipse reference time $T_0^e$, the
 orbital period $P_{\rm{orb\; 0}}$, and the orbital period derivative
 $\dot{P}_{\rm{orb}}$ obtained by PA00, BU10, BA10, and JI11 in
 Table \ref{Tab1}.

 In this work, we comprehensively examine both X-ray and optical/UV
 eclipse arrival times to give the most updated ephemeris of
 X1822-371, adding to the eclipse arrival times reported by BU10 the
 one obtained from a Suzaku observation performed in
 2006.  We also include a data point from a Ginga observation
 performed in 1989, and a data point from a ROSAT observation
 performed in 1992.  We critically examine the discrepancies that have
 emerged in calculating the orbital ephemeris in previous papers and,
 finally, show the ephemeris of the X1822-371 by combining the optical/UV
 and X-ray data-sets.

\begin{table*}[ht]
  \caption{Journal of the ephemerides of X1822-371 discussed in this work.}
\label{Tab1}      
\begin{center}
\begin{tabular}{l r r r r}          
\hline\hline                        
 Parameters  &
\cite{parmar2000}&\cite{Burderi_2010}&
 \cite{Bayless2010}&\cite{Ji2011}\\
\hline

$T_0^e$ (MJD$_{\odot}$) &  

45614.80964(15)&
45614.80948(14)&
45614.81166(74)&
45614.80927(25)\\

$P_{\rm{orb\; 0}}$ (s)&   
20054.1990(43)&
20054.2056(22)&
20054.1866(69)&
20054.2181(41)\\

$\dot{P}_{\rm{orb}}$ ($\times 10^{-10}$ s/s) &
1.78(20) &
1.499(71)&
2.12(18)&
0.827(95)\\

$\chi^2/(d.o.f.)$&
21.4/16&
38.69/25&
70.04/32&
35.99/19
\\

\hline                                             
\end{tabular}
\end{center}

{\small \sc Note} {\footnotesize---Uncertainties are at the  68\% 
  c. l. for a single parameter.  We show the reference time $T_0^e$ of the 
  eclipse arrival times in units of  MJD, the orbital period 
  $P_{\rm{orb\; 0}}$ in units of seconds calculated 
  at $T_0^e$,  the derivative of the orbital period  $\dot{P}_{\rm{orb}}$ in units of s/s, and finally the $\chi^2/(d.o.f.)$ obtained fitting the eclipse arrival times with a quadratic function.}  
\end{table*}

 \section{Suzaku observation}

 Suzaku observed X1822-371 on 2006 October 2 with an elapsed time of
 88 ks, the start and stop times of the observation corresponding to
 54010.48 and 54011.50 MJD, respectively.  Both the X-ray Imaging
 Spectrometer \citep[0.2-12 keV, XIS;][]{Koyama2007} and the Hard
 X-ray Detector \citep[10-600 keV, HXD;][]{Takahashi2007} instruments
 were used during these observations. In this work, we used only the
 XIS data.  There are four XIS detectors, numbered  0 to 3. The
 XIS0, XIS2, and XIS3 detectors use front-illuminated CCDs and have
 very similar responses, while XIS1 uses a back-illuminated CCD.

 We reprocessed the observation using the aepipeline tool included in
 the Suzaku FTOOLS Version 16 applying the latest calibration
 available as of 2011 March. During the observation, XIS0 and XIS1 were
 used adopting the quarter window option (frame time 2 s), while XIS2 and
 XIS3 worked in full window (frame time 8 s) mode.  We barycentred the XIS
 data using the SUZAKU tool aebarycen and adopting as the best
 estimate of the source coordinates those derived from the 2008
 Chandra observations (RA: 18 25 46.81, DEC: -37 06 18.5, uncertainty:
 0.6\arcsec).

\begin{figure}[ht]
\includegraphics[height=8cm,angle=0]{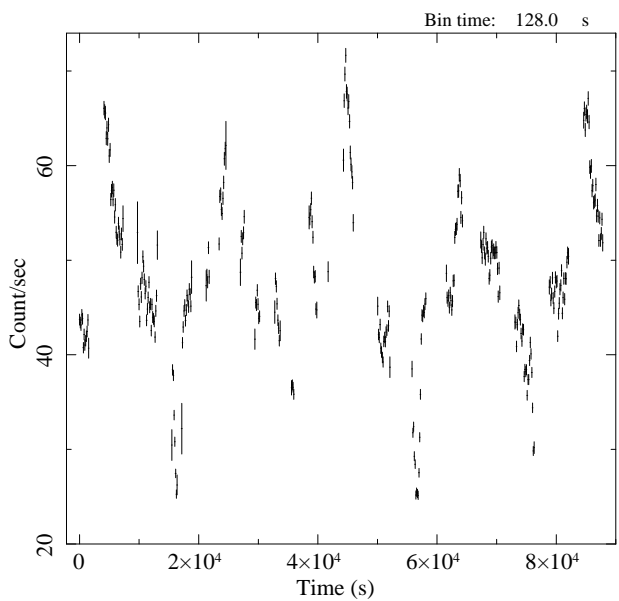}
\caption{Combined XIS light curve of X1822-371  in the 1-10 keV
  energy band. The adopted bin time is 128 s.}
\label{Fig1}
\end{figure}

 We extracted the four XIS light curves in the 1-10 keV energy band
 selecting a circular region centred on the source. We adopted a
 radius of 130 pixels for XIS0 and XIS1 and 160 pixels for XIS2 and
 XIS3.  The four light curves are quite similar and enclose 
four orbital periods of X1822-371, thus we used the FTOOL lcmath to
 combine the four XIS light curves. The combined XIS light curve is
 shown in Fig. \ref{Fig1} adopting a bin time of 128 s.
 
\begin{table}[ht]
  \caption{Journal of all the available X-ray eclipse times.}

\label{Tab_times}      
{\small
\begin{center}
\begin{tabular}{l c c c c}          
\hline\hline                        
Eclipse Time (JD$_{\odot}$)  &
Error&Cycle&
 Ref.& Satellite\\
\hline

2443413.5272 &  0.0046 & -9486 & 1 &HEAO-1 Scan \\
2443591.5521 &  0.0046 & -8719& 1  &HEAO-1 Scan \\
2443776.5459 &  0.0012 & -7922& 1& HEAO-1 Point\\
2443778.4065 &  0.0046 & -7914& 1&HEAO-1 Scan \\
2443969.4247 &0.0069   &-7091& 2 &Einstein\\
2444133.5277 &  0.0030 & -6384& 1 & Einstein \\
2445580.4932 &  0.0005 & -150 & 1 & EXOSAT\\
2445615.30940 & 0.00038& 0 &1& EXOSAT\\
2445963.00914 & 0.00033&1498&1& EXOSAT\\
2445963.24046 & 0.00030& 1499&1& EXOSAT\\
2445963.47254 & 0.00034& 1500&1& EXOSAT\\
2446191.63643 &0.00031&2483&1& EXOSAT\\
 2446191.86768& 0.00033&2484&1& EXOSAT\\
2446192.10008& 0.00029&2485&1& EXOSAT\\
2447760.22900& 0.00030&9241&1&Ginga\\
2448692.84396& 0.00070&13259 &1&ROSAT\\
2449268.00984& 0.00040&15737&3&ASCA\\
 2450353.35425& 0.00035&20413&3&ASCA\\
2450353.58728& 0.00023&20414&3&RXTE\\
2450701.51870 & 0.00120&21913&3&BeppoSAX\\
2450992.58580 &0.00230&23167&4&RXTE\\
2451780.13170& 0.00190&26560&4&Chandra\\
2451975.56934& 0.00056&27402&4&XMM-Newton\\
2451975.56935& 0.00031&27402&4&RXTE\\
2452432.59458& 0.00030&29371 &4&RXTE\\ 
 2452488.53300& 0.00038&29612 &4&RXTE\\ 
2452519.63569& 0.00085&29746&4&RXTE\\ 
2452882.65470& 0.00037&31310&4&RXTE\\ 
2454011.17300& 0.00090&36172&5&Suzaku\\ 
2454607.69592& 0.00056&38742&4&Chandra\\

\hline                                             
\end{tabular}
\end{center}}

{\small \sc Note} {\footnotesize---References. (1) \citealt{hellier_smale_94},
  (2) \citealt{Hellier_mason1989}, (3) \citealt{parmar2000}, 
(4) \citealt{Burderi_2010}, (5) 
this work. The number of cycles for each eclipse time is discussed
in sec. \ref{updat_sec}.}  
\end{table}

During only  the third orbital passage of X1822-371, the eclipse was
fully covered by Suzaku at a time of 55,000 s from the start time.  To
estimate the eclipse arrival time, we folded the combined XIS light
curve, adopting the ephemeris reported by BU10 and a bin time of 128
s. We fitted the orbital light curve to derive eclipse arrival times by
adopting the same procedure described in BU10, obtaining 
an eclipse time passage at  $54,010.6730 \pm 0.0009$
MJD$_{\odot}$ with an associated error at the 68\% confidence level.

\section{The ephemeris of X1822-371}
For clarity's sake, we show in Table \ref{Tab_times} the X-ray eclipse
arrival times that we used to update the X-ray ephemeris of X1822-371.
Most of these data points were included in the timing analysis
of BU10. To their data set, we added eclipse arrival times from Ginga
(1989), ROSAT (1992), and, most importantly, Suzaku (2006). The
  ephemerides showed in Table \ref{Tab1} and in the analysis now described
  are given in barycentric dynamical time. We  note that the RXTE
arrival times from 1998 to 2003 reported in Tab. 1 of BU10 (except for
the second point corresponding to cycle 23,167) are not the eclipse
arrival times, as erroneously stated, but the times of passage through
the ascending node (which differs from the eclipse time by
$P_{\rm{orb}}/4$). Nevertheless, the corresponding RXTE time delays
were correctly shown in Fig. 1 of BU10 and correctly used to derive
the orbital ephemeris, which are therefore  unaffected by this
mistake. The correct RXTE eclipse arrival times are shown in our Table
\ref{Tab_times}. The  X-ray ephemeris of X1822-371 reported by
BU10 and JI11 show a large discrepancy in the quadratic term by almost
a factor of two.  JI11 suggested that the discrepancy in the time delay
associated with the last two Chandra observations is caused by  
 BU10  not folding the Chandra light curves to estimate the
eclipse arrival time, which instead was done by BU10.  To
understand the reason for this discrepancy, as a first step we tried to
reproduce the results of JI11 by using the same data they used in
their analysis. These consist of a total of 22 eclipse-times, which are,
respectively, those given by \cite{hellier_smale_94}, PA00, and three
eclipse arrival times obtained from three Chandra observations
corresponding to obsID 671, 9076, and 9058 derived by JI11 (see Tab. 2
in their paper).

\begin{figure}[ht]
\includegraphics[height=8.5cm,angle=-90]{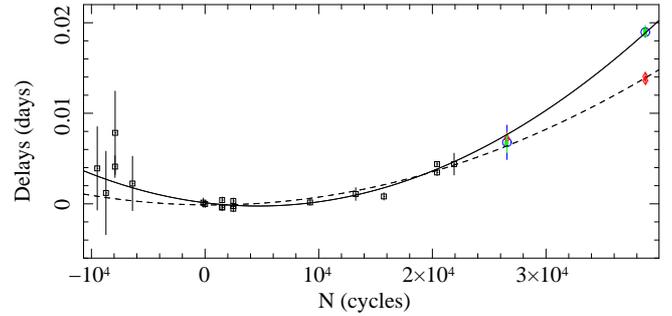}
\caption{Time delays of the eclipse arrival times with respect to the
  linear ephemeris of \cite{hellier_smale_94}. The black open squares
  correspond to the eclipse arrival times shown by
  \cite{hellier_smale_94} and PA00, the red diamonds to the Chandra
  eclipse arrival times given by JI11, the green filled squares to the
  Chandra eclipse arrival times reported in this paper (see
  Table \ref{Tab_chandra}), the blue open circles to the Chandra eclipse
  arrival times reported by BU10. The dashed and solid lines are the
  best-fit quadratic curve obtained by JI11 and in this paper (see text),
  respectively.}
\label{Fig_comp}
\end{figure}

We found the corresponding time delays following their procedure,
namely we determined the time delays with respect to the best-fit
linear ephemeris shown by \cite{hellier_smale_94}, that is
 $$
T_{ecl}=2445615.30942(14) {\rm JD_{\odot}}+0.232109017(33) N,
$$
and fitted the time delays with a quadratic function.  We obtained
best-fit values consistent with the ones reported in JI11.  We
showed in Fig. \ref{Fig_comp} the time delays in units of days
associated with the eclipse arrival times shown by
\cite{hellier_smale_94} and PA00 with black open squares. The delay
times associated with the Chandra eclipse arrival times showed by JI11
were plotted using red diamonds.  The dashed line corresponds to the
quadratic best-fit curve given by JI11.

After  establishing the reproducibility of the parameter's
estimates of JI11, we explored the cause of the discrepancy in
the fitting results given by BU10 extracting the eclipse arrival times 
from each Chandra barycentred and folded light curve 
(obsIDs 671, 9076, and 9858). 
We show the Chandra eclipse arrival times in Table \ref{Tab_chandra}.
The corresponding delays were estimated as described above. 
\begin{table}[ht]
  \caption{Eclipse arrival times of the three Chandra observations.}

\label{Tab_chandra}      
{\small
\begin{center}
\begin{tabular}{l  l l l l}          
\hline\hline                          
ObsID &
Eclipse Time (MJD$_{\odot}$)  &
Cycle &Delay (days)\\
\hline    
671& 51779.8638(11) & 26561& 0.0067(11)\\
9076& 54607.19610(56) & 38742& 0.01914(56) \\
9858& 54609.74890(31) & 38753 & 0.01875(31) \\ 
\hline
\hline                                             
\end{tabular}
\end{center}}

{\small \sc Note} {\footnotesize---Column 1: Chandra obsIDs. Column 2: 
our estimation of the eclipse arrival time. Column 3: the corresponding cycle
with respect to the linear ephemeris given by  \cite{hellier_smale_94}. Column 
4: the corresponding delay in units of days. The errors are at the 68\% c.l.}  
\end{table}

We found that the discrepancies of the eclipse arrival times
between our analysis and JI11's are  $-40 \pm 140$ s, $440 \pm 70$ s,
 and $440 \pm 50$ s for obsIDs 671, 9076,  and 9858, respectively.
  Since the errors are at the  68\% c. l., the eclipse arrival
times corresponding to the obsIDs 9076 and 9858 are not compatible.
We show our Chandra delays with green filled squares in
Fig. \ref{Fig_comp}.  Only two eclipse times were derived by BU10 from
the Chandra observations, because light curves of obsID 9076 and
obsID 9058 were combined to obtain a single folded light-curve and a
single eclipse-time passage, with a smaller uncertainty, since the
observations were sufficiently close in time to each other. We show
the two corresponding time delays with blue open circles in
Fig. \ref{Fig_comp}. We note that our delays and those given by BU10
are widely compatible.
  
Fitting the
time delays corresponding to the eclipse arrival times given by 
\cite{hellier_smale_94}, PA00, and our three eclipse arrival times 
reported in Table \ref{Tab_chandra}, for a total of 22 data points,  
we obtained
\begin{multline} 
\label{ji}
T_{ecl}=45614.80954(14) {\rm MJD_{\odot}}+0.2321088628(21)N\\
+1.648(72)\times 10^{-11}N^2,
\end{multline} 
with a $\chi^2/(d.o.f.)=25.6/19$ and the errors are at the  68\%
confidence level, the uncertainties in the parameters have been
  scaled by a factor $\sqrt{\chi^2_{red}}$ to take into account a
  $\chi^2_{red}$ of the best-fit model larger than 1.  We note
that the quadratic term is larger than that shown by JI11.  The
corresponding $P_{\rm{orb\; 0}}$ and $\dot{P}_{\rm{orb}}$ are
20054.20574(17) s and $1.420(63) \times 10^{-10}$ s/s, respectively.
These values are compatible within one $\sigma$ with the ones given by
BU10 (see Table \ref{Tab1}).


\subsection{Updated X-ray ephemeris of X1822-371}
\label{updat_sec}
As a first step, we found the X-ray ephemeris of X1822-371 using the
eclipse arrival times adopted by BU10 excluding the Chandra eclipse
arrival times and including the eclipse arrival times taken with
Ginga, ROSAT, and Suzaku (see Tab. \ref{Tab_times}) for a total of 28
available data points. The Suzaku data-point is very important in this
respect, since it was taken in 2006 and therefore precedes the last
Chandra data-points taken in 2008. This is very important to fill the
time gap between the last RXTE arrival time taken in 2003 and the most
recent Chandra observation taken in 2008, and therefore gives us the
opportunity to discriminate more clearly between the Chandra eclipse
arrival time as reported by JI11 and our measurement (which is
compatible with the one reported by BU10).

We found the delays of the eclipse
arrival times by subtracting from our measurements the eclipse arrival
times predicted by a constant orbital period model adopting the
orbital period, $P_{\rm{orb\; 0}}$, and the reference time, $T_0^e$,
given by PA00. The time delays were plotted versus the
orbital cycle number N. The integer N is the exact number of orbital
cycles elapsed since $T_0^e$; the cycle number N corresponding to each
eclipse arrival time is shown in column 3 of
Table \ref{Tab_times}. We then  fitted the time delays using a parabolic
function obtaining a $\chi^2/(d.o.f.)$ of 33.63/25. We found that $T_0^e =
45614.80959(16)$ MJD$_{\odot}$, $P_{\rm{orb\; 0}} = 20054.2020(28)$ s,
and $\dot{P}_{\rm{orb}} = 1.626(90) \times 10^{-10}$ s/s with the
associated errors at the 68\% confidence level. All  these values are
compatible within one $\sigma$ with those given by BU10
(see column 3 in Table \ref{Tab1}), this suggests that the
Chandra eclipse arrival times given by BU10 are
in agreement with all the previous points.

To update the X-ray ephemeris of X1822-371, we then included the
Chandra eclipse arrival times given by BU10 in our data set for a
total of 30 available data points. We found the corresponding delays
and cycle numbers as described above.
\begin{figure}[ht]
{\includegraphics[height=8.5cm,angle=-90]{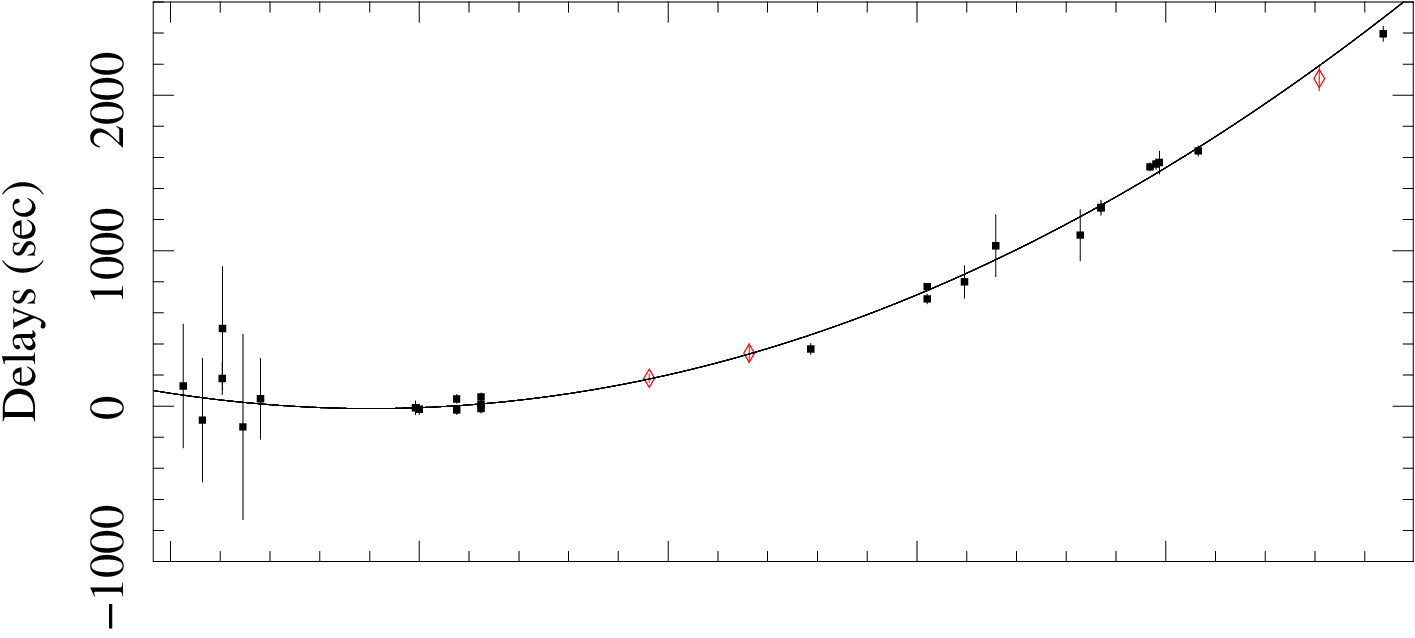}\\
 \includegraphics[height=8.5cm,angle=-90]{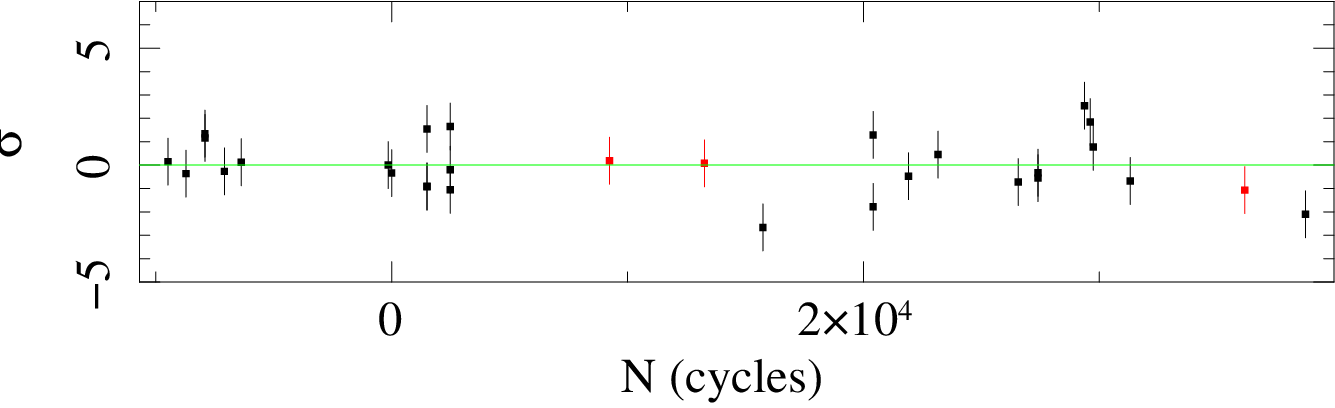}
}
\caption{Upper panel: Eclipse time delays with respect to a constant
  orbital period model plotted versus  the orbital cycle for all the
  available X-ray eclipse time measures together with the best-fit
  parabola. Lower panel: residuals in units of $\sigma$ with respect to
  the best-fit parabola. The black full squares points are from
  BU10, the red diamonds are the data added in this
  work.}
\label{Fig3}
\end{figure}
The time delays are shown in the upper panel of Fig. \ref{Fig3}.  We
 plotted the time delays used by BU10 with black full squares,
while the time delays added in this work and corresponding to the
Ginga, ROSAT, and Suzaku eclipse times are shown with red diamonds.
We then fitted the time delays using a parabolic function, resulting
in the  ephemeris
\begin{multline} 
\label{eq_x}
T_{ecl}=45614.80953(16) {\rm MJD_{\odot}}+0.232108853(30)N\\
+1.757(93)\times 10^{-11}N^2,       
\end{multline} 
where the associated errors are at the 68\% confidence level.  We obtained
a $\chi^2/(d.o.f.)$ of 41.2/27, the best-fit curve is shown with a
solid line in Fig.  \ref{Fig3}. We show the residuals in units of $\sigma$ 
in the lower panel of Fig. \ref{Fig3} and report the obtained values of 
$T_0^e$, $P_{\rm{orb\; 0}}$, and $\dot{P}_{\rm{orb}}$ in the second column of 
Table \ref{Tab2}. 

We found that the derivative of the orbital period,
$\dot{P}_{\rm{orb}}$, is $1.514(80) \times 10^{-11}$ s/s, 
compatible with the value of $1.499(71) \times 10^{-11}$ s/s estimated
by BU10.

\begin{table}[ht]
  \caption{Updated X-ray and optical/UV ephemeris of X1822-371.}
\label{Tab2}      
{\tiny                               
\begin{center}
\begin{tabular}{l c c}          
\hline\hline                        
 Parameters  &
X-ray& 
 Optical/UV\\
\hline

$T_{0}^e$ (MJD$_{\odot}$) &  

45614.80953(16)&
45614.8116(11)\\

$P_{\rm{orb\; 0}}$ (s)&   
20054.2049(26)&
20054.188(10)\\

$\dot{P}_{\rm{orb}}$ &
 1.514(80) &
2.10(26)\\

$\chi^2/(d.o.f.)$&
41.2/27&
71.38/33\\

\hline  
\end{tabular}
\end{center}
}
{\small \sc Note}{ \footnotesize---Uncertainties are at the  68\% 
  c. l. for a single parameter.  The parameters are defined as in
  Table \ref{Tab1}, the derivative of the orbital period is in units of 
  $10^{-10}$ s/s.
  The updated values of $T_0^e$, $P_{\rm{orb\; 0}}$, and $\dot{P}_{\rm{orb}}$ 
  using  X-ray data and  optical/UV data are shown in Cols. 2 and  3, 
  respectively. } 
\end{table}

\subsection{Updated optical/UV ephemeris of X1822-371}
 BA10 used 35 optical/UV eclipse arrival times shown in
Tab. 1 of their paper to find the best-fit optical/UV ephemeris of
X1822-371 given by
\begin{multline} 
\label{eq_opt_uv}
T_{ecl}=45614.81166(74) {\rm MJD}+0.232108641(80)N\\
+2.46(21)\times 10^{-11}N^2,
\end{multline} 
where the errors are at the 68\% confidence level (Bayless, private
communication).  We added to their data the optical eclipse arrival
time $2,443,629.841 \pm 0.013$ JD$_{\odot}$ given by
\cite{Hellier_mason1989} and not included in BA10.

\begin{figure}[ht]
{\includegraphics[height=8.5cm,angle=-90]{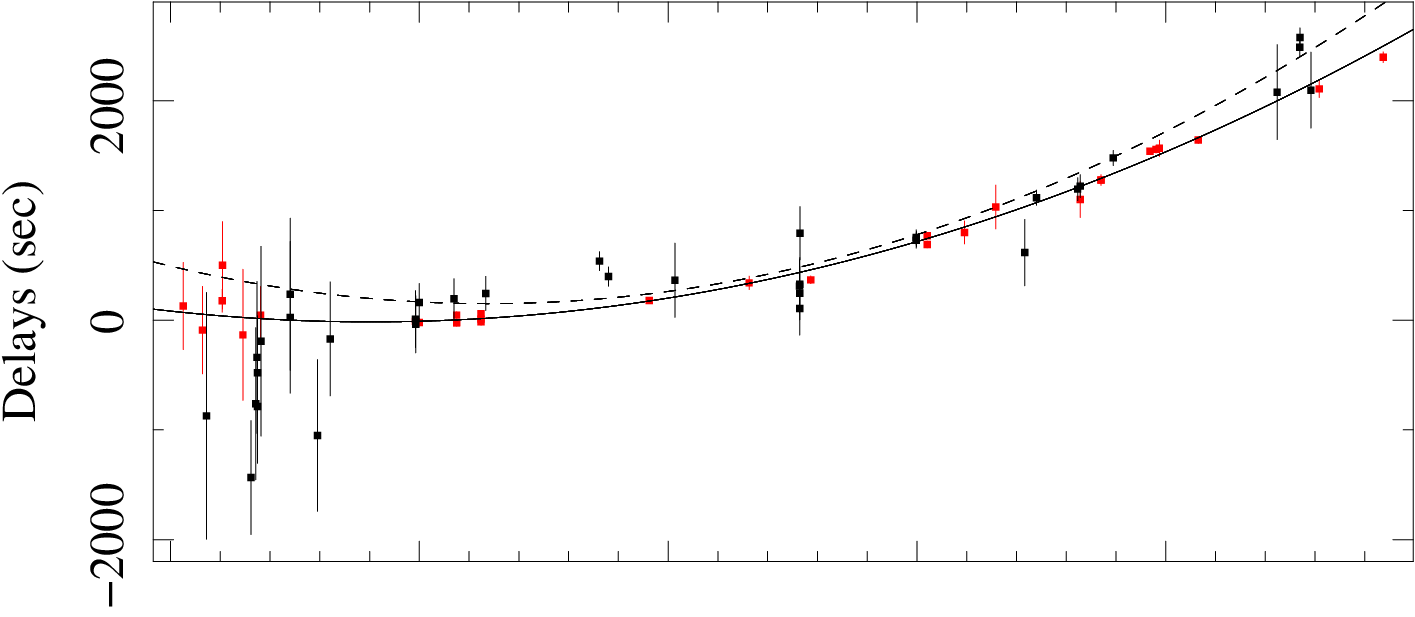}\\
\includegraphics[height=8.5cm,angle=-90]{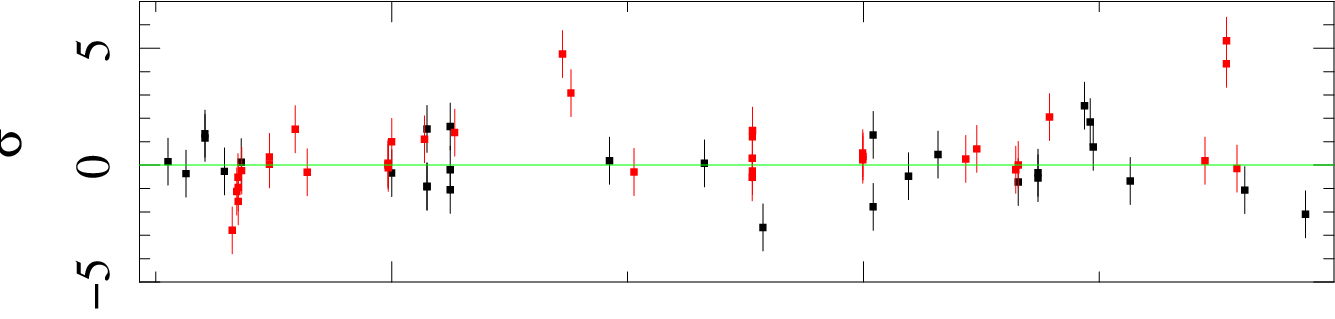}\\
\includegraphics[height=8.5cm,angle=-90]{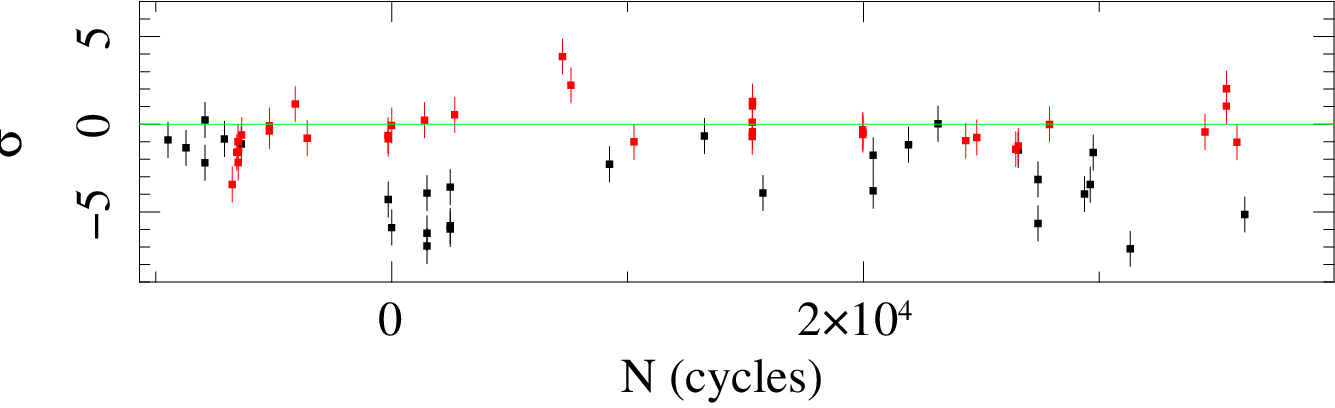}
}
\caption{Top panel: the optical/UV (black filled squares) and X-ray
  (red filled squares) time delays. The dashed and solid lines
  correspond to the optical/UV and X-ray best-fit parabolic
  curve. Middle panel: residuals with respect to the X-ray best-fit
  parabolic curve.  Bottom panel: residuals with respect to the
  optical/UV best-fit parabolic curve.}
\label{Fig4}
\end{figure}

Using the 36 optical/UV data points and following the procedure described in
the previous section we found the corresponding time delays. Fitting
them with a parabola, we obtained the following optical/UV ephemeris
\begin{multline} 
\label{eq_opt_uv_mie}
T_{ecl}=45614.8116(11) {\rm MJD_{\odot}}+0.23210865(12)N\\
+2.44(31)\times 10^{-11}N^2,
\end{multline} 
with a $\chi^2/(d.o.f.)$ of $71.38/33$ and the errors are at the 68\%
confidence level.   The uncertainties have been
  scaled by a factor $\sqrt{\chi^2_{red}}$ to take into account a
  $\chi^2_{red}$ of the best-fit model larger than 1. This
  explains why the uncertainties in the optical/UV ephemeris we have
  shown are larger than the ephemeris shown by BA10.  The updated optical/UV
ephemeris are consistent with those given by BA10. We report the
corresponding values of $T_0^e$, $P_{\rm{orb\; 0}}$, and
$\dot{P}_{\rm{orb}}$ in the third column of Table \ref{Tab2}.  In the
upper panel of Fig. \ref{Fig4}, we show the time delays for each
eclipse arrival time of X1822-371 for the X-ray (red full squares) and
optical/UV bands (black full squares) for a total of 66 data points.  The
solid and dashed lines correspond to the best-fit parabolas
reproducing the X-ray and optical/UV ephemerides showed in
eqs. \ref{eq_x} and \ref{eq_opt_uv_mie}, respectively.

We compare the X-ray and optical/UV residuals with  the X-ray
best-fit parabola in Fig. \ref{Fig4} (middle panel).  Although we plot
the residuals for the best-fit parabola obtained from the
X-ray time delays, we note that almost all of the optical/UV data are
close to the best-fit curve.  The largest discrepancies are associated with
the last two optical eclipse times shown by \cite{Hellier_mason1989}
corresponding to orbital cycles 7,243 and 7,600 and  the
two UV eclipse arrival times obtained with HST and reported by
BA10; these last two data points are at orbital cycles 35,387
and 35,395, respectively. All the other optical/UV points are
within two $\sigma$ of the corresponding values of the best-fit X-ray
ephemeris.

In Fig. \ref{Fig4} (bottom panel), we show the X-ray and optical/UV
residuals with respect to  the optical/UV best-fit curve.  In this case, the
X-ray data are mainly below the best-fit optical/UV parabola.

\subsection{ Time-lag between optical/UV and X-ray eclipse times}
 To our knowledge, there are only two simultaneous X-ray and optical 
  observations of the eclipse of X1822-371 reported by 
  \cite{Hellier_mason1989} and \cite{Hellier90}; these authors 
   showed that the optical eclipse times lag 
  the X-ray eclipse times by $3.0 \pm 3.4$ min, and $180 \pm 50$ s, 
  respectively. 

  The optical eclipses are also wider than the X-ray eclipses; the
  different width suggests a different origin for the optical and
  X-ray eclipses, respectively.  \cite{Hellier_mason1989} proposed
  that the X-ray emission comes from an accretion disc corona (ADC) with
  a radius half of the outer accretion disc radius, while the optical
  emission is produced by a more extended disk structure. Furthermore,
  the optical eclipse lags the X-ray eclipse because of the asymmetric
  disk structure probably caused by the stream impact onto the outer
  accretion disk. \cite{Hellier_mason1989}, modelling the X-ray and
  optical light curves of X1822-371, found an optical eclipse time-lag
  of $\sim 0.01$ in units of orbital phase, corresponding to a
  time-lag of 200 s.  BA10 discussed  a marginally
  significant time-lag between the optical/UV and X-ray ephemeris of
  $100 \pm 65$ s and 122 s with respect to the X-ray ephemeris reported
  by PA00 and BU10, respectively.

 Since we have used an unprecedentedly large amount of optical/UV
  and X-ray eclipse times, we can estimate the average time-lag along
  50,000 orbital cycles with good accuracy.  We fitted simultaneously
  the X-ray and optical/UV time delays allowing the constant terms of each
  parabola free to vary and constraining the values of the linear and
  quadratic parameters of each parabola to the same value, since the
  orbital period of X1822-371 and its derivative  cannot depend on the
  considered waveband.
 
  Fitting the time delays, we obtained a large $\chi^2/(d.o.f.)$ of
  $124.04/62$ and found that the best-fit values of the linear and
  quadratic terms are $(3.7 \pm 2.8) \times 10^{-3}$ s and $(1.595 \pm
  0.086) \times 10^{-6}$ s, respectively. The constant terms are $121
  \pm 36$ s and $-6 \pm 16$ s for the optical/UV and X-ray data-sets,
  respectively.  Using these values, we obtained the
  ephemerides for the X-ray and optical/UV data
\begin{multline} 
\label{eq_opt_uv_x_forx}
T_{ecl_{X-ray}}=45614.80957(19) {\rm MJD_{\odot}}+0.232108828(32) N\\
+1.847(99)\times 10^{-11}N^2,
\end{multline} 
\begin{multline} 
\label{eq_opt_uv_x_foroptuv}
T_{ecl_{opt/UV}}=45614.81104(42) {\rm MJD_{\odot}}+0.232108828(32) N\\
+1.847(99)\times 10^{-11}N^2.
\end{multline} 

\begin{figure}[ht]
{\includegraphics[height=8.5cm,angle=-90]{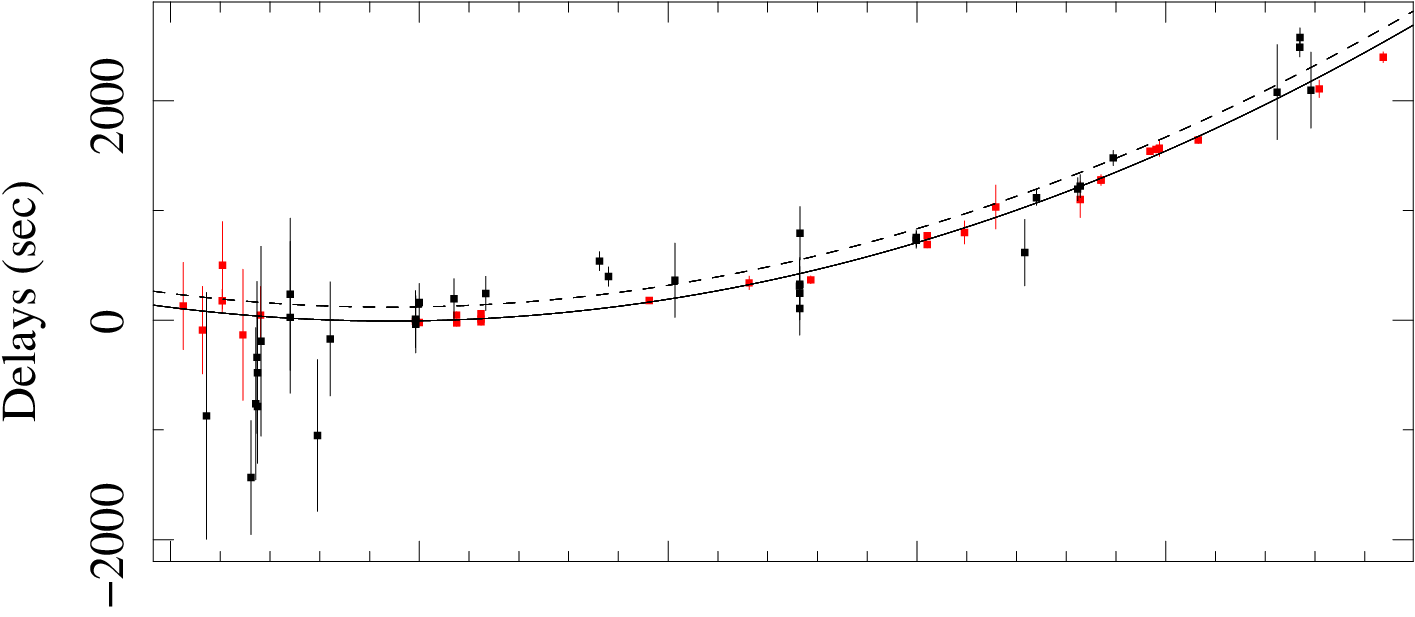}\\
\includegraphics[height=8.5cm,angle=-90]{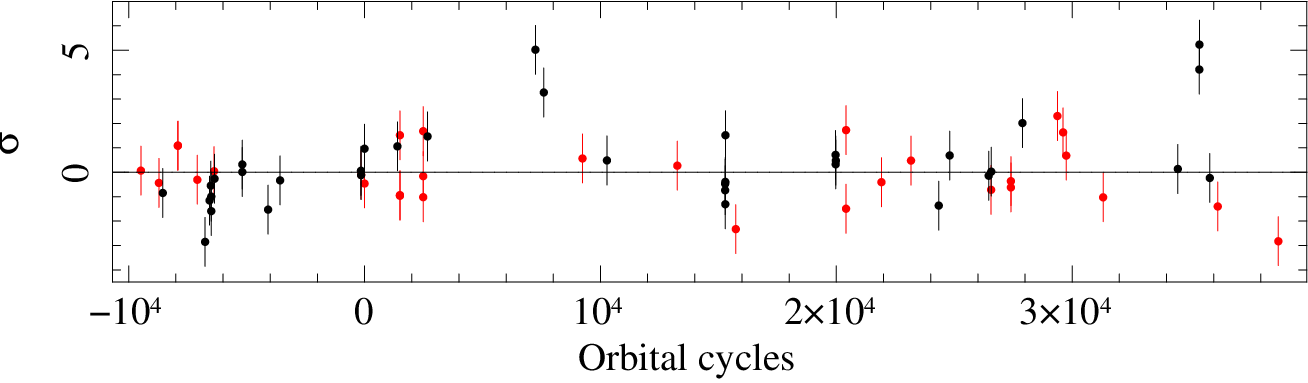}
}
\caption{Upper panel: the optical/UV (black filled squares) and X-ray
  (red filled squares) time delays fitted with two parabolas having
  the same linear and quadratic terms.  The solid and dashed parabolas
  correspond to the X-ray and optical/UV best-fit curves. Lower panel:
  Residuals in units of $\sigma$ with respect to the best-fit parabola 
  describing the X-ray ephemeris shown in eq. \ref{eq_opt_uv_x_forx}. }
\label{Fig5}
\end{figure}

 The corresponding orbital period derivative is  $1.591(86)
  \times 10^{-10}$ s/s, and the reference time $T_0$ is $45614.80957(19)$
  and $45614.81104(42)$ MJD$_{\odot}$ for X-ray and optical/UV data-sets,
  respectively; all the errors are at the 68\% c.l. For clarity's sake,
  the values of $T_0^e$, $P_{\rm{orb\; 0}}$, and $\dot{P}_{\rm{orb}}$
  are showed in  Table \ref{Tab7}.
  In the upper panel of Fig. \ref{Fig5}, we show the X-ray (red points)
  and the optical/UV (black points) time  delays; the dashed and solid
  curves are the optical/UV and X-ray best-fit parabolas,
  respectively.

  From our analysis, we found a time-lag of $127 \pm 52$ s, which is
  significant at a confidence level of 2.4 $\sigma$. In the bottom
  panel of Fig. \ref{Fig5}, we show the residuals of the X-ray (red
  points) and optical/UV (black points) delays with respect to the
  X-ray best-fit parabola.

\begin{table*}[ht]
  \caption{Ephemeris of X1822-371
fitting simultaneously X-ray  and optical/UV data.}
\label{Tab7}      
{                               
\begin{center}
\begin{tabular}{c c c c c}          
\hline\hline                        
   
$T_{0_{X-ray}}^e$  &
$T_{0_{optical/UV}}^e$ &  
$P_{\rm{orb\; 0}}$ &
$\dot{P}_{\rm{orb}}$ &
$\chi^2/(d.o.f.)$\\
(MJD$_{\odot}$) &(MJD$_{\odot}$) & (s) & $(\times 10^{-10})$ s/s&\\ 
  \hline             


45614.80957(19)&
45614.81104(42)&
20054.2027(28)&
 1.591(86) &
124.04/62\\

\hline  
\hline  
\end{tabular}
\end{center}
}
{\small \sc Note}{ \footnotesize---Uncertainties are at  68\% 
  c. l. for a single parameter.  The parameters are defined as in
  Table \ref{Tab1}. The uncertainties in the parameters have been 
  scaled by a factor $\sqrt{\chi^2_{red}}$ to take
  into account a $\chi^2_{red}$ of the best-fit model larger than 1.} 
\end{table*}

 The values of the optical/UV time-lags with respect to the
  best-fit parabola describing the X-ray ephemeris given by
  eq. \ref{eq_opt_uv_x_forx} are shown in Table \ref{Tab_opt}.  We note
  that the largest optical time-lags are associated with the last two
  optical eclipse times shown by \cite{Hellier_mason1989}
  corresponding to orbital cycles 7,243 and 7,600 and with the recent
  two UV eclipse arrival times obtained with HST and reported by BA10;
  the corresponding time-lags are $433 \pm 86$ s, $282 \pm 86$ s, $364
  \pm 86$ s, and $452 \pm 86$ s, respectively.  These values are
  larger than 200 s,  in disagreement with that  predicted by
  \cite{Hellier_mason1989} modelling the X-ray and optical light
  curves of X1822-371.

 \begin{table*}[ht]
  \caption{Journal of the optical/UV time-lags}
\label{Tab_opt}      
{\small
\begin{center}
\begin{tabular}{c r r r}          
\hline\hline                        
Eclipse Time 
& Time-lag & residuals best-fit 1& residuals best-fit 2\\
(JD$_{\odot}$-2,400,000) & (s) & (s) &(s)\\
\hline
\hline 

$43629.8410 \pm 0.0130$&$-951 \pm  1123$ & $-938 \pm 1123$ & $-1278 \pm 1123$\\ 
$44044.8450 \pm 0.0060$&$ -1477 \pm 518$&$-1444\pm  518$&$-1395\pm  518$\\
$44090.1140 \pm 0.0080$&$ -800 \pm 691$&$-767 \pm  691$&$-644 \pm  691$\\
$44101.0280 \pm 0.0080$&$ -377 \pm 691$&$-344 \pm  691$&$-241 \pm  691$\\
$44105.6650 \pm 0.0060$&$ -824 \pm 518$&$-791 \pm 518$&$-701 \pm 518$\\ 
$44106.5970 \pm 0.0060$&$ -516 \pm 518$&$-483 \pm 518$&$-396 \pm 518$\\ 
$44137.9350 \pm 0.0100$&$ -227 \pm 864$&$-195 \pm  864$&$-252 \pm  864$\\ 
$44411.1320 \pm 0.0080$&$ 218  \pm 691$&$242 \pm  691$&$247 \pm  691$\\ 
$44412.0580 \pm 0.0080$&$ 8 \pm  691  $&$32 \pm 691$&$32 \pm 691$\\ 
$44664.8120 \pm 0.0080$&$ -1057 \pm 691$&$-1051 \pm  691$&$-916 \pm  691$\\ 
$44783.8940 \pm 0.0060$&$ -172 \pm 518$&$-179 \pm 518$&$-542 \pm 518$\\ 
$45579.5650 \pm 0.0030$&$  16 \pm 259$&$-114\pm 259$&$-163\pm 259$\\
$45580.7250 \pm 0.0030$&$ -31 \pm 259$&$-162 \pm 259$&$-216 \pm 259$\\ 
$45615.3115 \pm 0.0020$&$ 166 \pm 173$&$ -30 \pm 173$&$ -173 \pm 173$\\ 
$45937.7110 \pm 0.0021$&$ 192 \pm 181$&$-4 \pm  181$&$-160 \pm  181$\\ 
$46234.5787 \pm 0.0018$&$ 228 \pm 156$&$-20 \pm 156$&$-83 \pm 156$\\ 
$47296.4798 \pm 0.0010$&$ 433 \pm 86$&$80 \pm 86$&$158 \pm 86$\\ 
$47379.3410 \pm 0.0010$&$ 282 \pm 86$&$-72 \pm  86$&$25 \pm  86$\\ 
$47999.7674 \pm 0.0039$&$ 163 \pm 337$&$-160 \pm 337$&$209 \pm 337$\\ 
$49163.0960 \pm 0.0028$&$ -114 \pm 242$&$-253 \pm  242$&$37 \pm  242$\\ 
$49164.0237 \pm 0.0028$&$ -178 \pm 242$&$-317 \pm 242$&$-25 \pm 242$\\ 
$49164.2542 \pm 0.0028$&$ -317 \pm 242$&$-456 \pm 242$&$-164 \pm 242$\\ 
$49165.1852 \pm 0.0028$&$ -96  \pm 242$&$-234 \pm 242$&$59 \pm 242$\\ 
$49166.1190 \pm 0.0028$&$  368 \pm 242$&$229 \pm 242$&$524 \pm 242$\\ 
$50250.5308 \pm 0.0008$&$ 49 \pm  69$&$65 \pm  69$&$22 \pm  69$\\ 
$50250.7626 \pm 0.0008$&$ 22 \pm 69$&$38 \pm 69$&$-3 \pm 69$\\ 
$50252.6196 \pm 0.0008$&$ 33 \pm  69$&$49 \pm 69$&$18 \pm 69$\\ 
$51264.8446 \pm 0.0035$&$ -413 \pm 302$&$-409 \pm  302$&$-713 \pm  302$\\ 
$51373.7094 \pm 0.0008$&$ 47 \pm 69$&$40 \pm 69$&$-32 \pm 69$\\ 
$51756.4577 \pm 0.0012$&$  -15 \pm 104$&$-74 \pm 104$&$78 \pm 104$\\ 
$51782.4542 \pm 0.0012$&$  3  \pm 104$&$-60 \pm  104$&$-34 \pm  104$\\ 
$52089.7692 \pm 0.0008$&$  139 \pm 69$&$24 \pm 69$&$-40 \pm 69$\\ 
$53618.6767 \pm 0.0050$&$  58 \pm 432$&$-285 \pm 432$&$-134 \pm 432$\\ 
$53828.9720 \pm 0.0010$&$  364 \pm 86$&$11 \pm  86$&$11 \pm  86$\\ 
$53830.8299 \pm 0.0010$&$  452 \pm 86$&$ 99 \pm 86$&$ 95 \pm 86$\\ 
$53932.7201 \pm 0.0040$&$ -80 \pm  346$&$-434 \pm 346$&$-93 \pm 346$ \\

\hline
\hline 

\end{tabular}
\end{center}}

{\small \sc Note} {\footnotesize---The optical/UV  eclipse times (1st column),
  the corresponding time-lags (2nd column) with respect to  the
  best-fit parabola describing the X-ray ephemeris showed in
  eq. \ref{eq_opt_uv_x_forx},  the time-lags after removal of
  the sinusoidal modulations (3rd and 4th column, respectively).
  The errors are at the 68 \% c.l.}  
\end{table*}
\begin{figure}[h]
{\includegraphics[height=8.3cm,angle=-90]{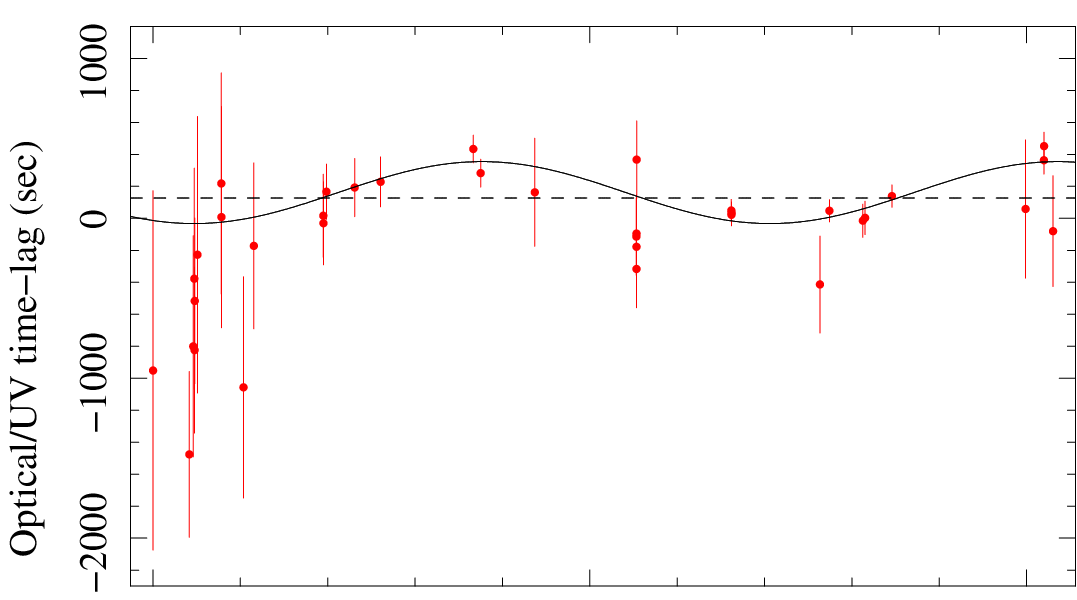}\\
\includegraphics[height=8.3cm,angle=-90]{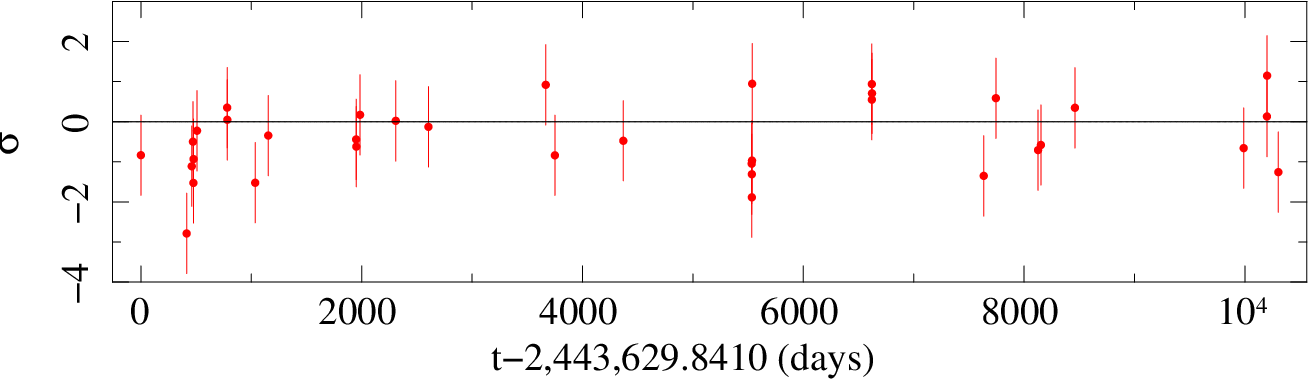}
}
\caption{Upper panel: the optical/UV (red filled squares) time-lags
  fitted with a constant (dashed line) and a sinusoidal function
  $f(t)$ (solid line) with a period of 18 yrs (see 
  text). Lower panel: Residuals in units of $\sigma$ with respect to
  the best-fit sinusoidal function.  }
\label{pino}
\end{figure} 
\begin{table}[ht]
  \caption{Best-fit parameters of the sinusoidal modulation fitting
 the optical/UV time-lags.}
\label{Tab_sin}      
{                               
\begin{center}
\begin{tabular}{l r r}          
\hline\hline                        

Parameters  & best-fit 1 & best-fit 2 \\ 

  \hline             
 
A (s) & $161 \pm 24$ & $105 \pm 24$ \\
B (s) & $194 \pm 25$ & $-267 \pm 43$ \\
P (days) & $6593 \pm 452$ & $283.1 \pm 0.6$ \\
t$_0$ (days) & $-1180 \pm 481$ &$239 \pm 16$ \\ 
 $\chi^2$(dof) & 34.52(32) & 35.57(32) \\

\hline  
\hline  
\end{tabular}
\end{center}
}
{\small \sc Note}{ \footnotesize---Uncertainties are at  the 68\% 
  c. l. for a single parameter.  In the second and third columns,
we show the best-fit parameters for the modulation of 18 years and
283 days, respectively. } 
\end{table}

 We fitted the optical/UV time-lags with a constant obtaining
  a large $\chi^2(d.o.f.)$ of $81.25(35)$; the constant value was 127
  s, which is similar to the averaged time-lag previously discussed.
  In Fig. \ref{pino} (top panel), we show the optical/UV time-lags as a
  function of time in units of days, the dashed line being the
  constant function. We  note that if we remove the four optical/UV
  points mentioned above, we find no significant time-lag between the 
  optical/UV and X-ray eclipse times, with a best-fit value of $31 \pm
  46$ s.

  Because we found a large value of $\chi^2_{red}$, after a visual
  inspection of the fit residuals, we decided to fit the time-lags
  with the function $f(t)=A-B \sin{\left[2\pi/P(t-t_0)\right]}$. In
  this case, we largely improved the fit for two different sets of
  parameters. For the first set, we obtained a $\chi^2(d.o.f.) =
  34.52(32)$ and a probability of chance improvement with respect to
  the fit with a constant of $4.03 \times 10^{-6}$. The values of the
  best-fit parameters are $A=161 \pm 24$ s, $B= 194 \pm 29$ s, $P=6593
  \pm 452 {\; \rm d}$ $(18.1 \pm 1.2$ yr), and $t_0 = -1180 \pm 481$ d,
  the errors being at the 68\% c.l..  For the second set of parameters, we
  obtained a $\chi^2(d.o.f.) = 35.57(32)$ and a probability of chance
  improvement with respect to the fit with a constant of $6.45 \times
  10^{-6}$; in this case, $A=105 \pm 24$ s, $B=-267 \pm 43$ s, $P=283.1
  \pm 0.6$ d, and $t_0 = 239 \pm 16$ d, the errors being at the 68\%
  c.l.. The best-fit values of both the fits are shown in
  Table \ref{Tab_sin}.

\begin{figure}[h]
{\includegraphics[height=8.3cm,angle=-90]{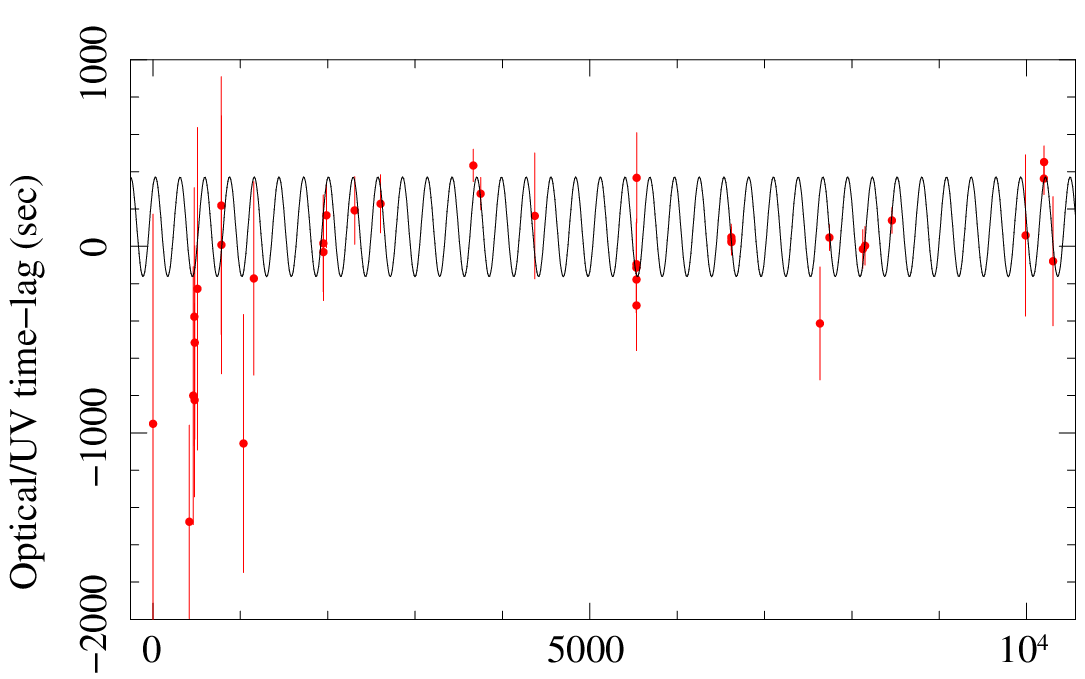}\\
\includegraphics[height=8.3cm,angle=-90]{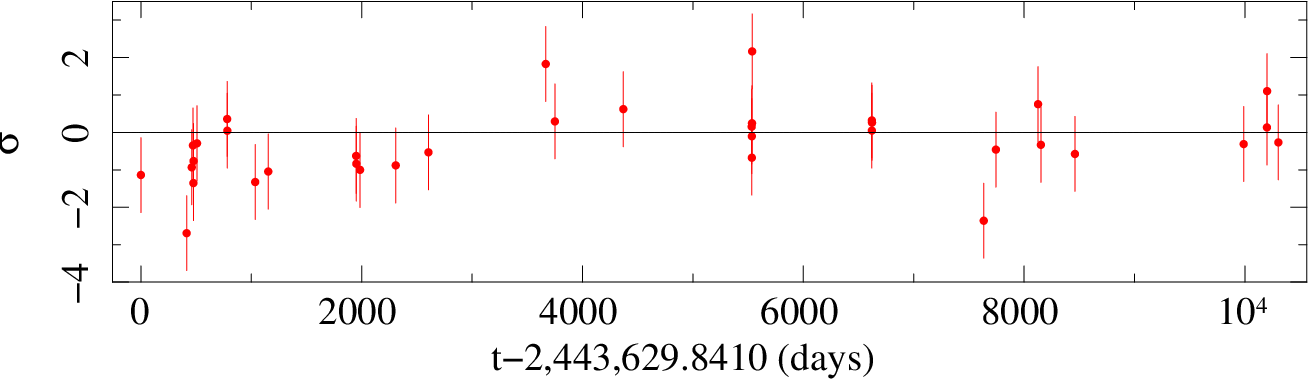}
}
\caption{Upper panel: the optical/UV (red filled squares) time-lags
  fitted with a  sinusoidal function $f(t)$ (solid line) with a
  period of 283 days (see text). Lower panel: Residuals in units of
  $\sigma$ with respect to the best-fit sinusoidal function.  }
\label{pino2}
\end{figure} 

 The best-fit curve corresponding to the first set of parameters
  (hereafter best-fit 1) is shown in Fig. \ref{pino} (top panel) with
  a solid line. In the bottom panel of Fig.  \ref{pino}, we show the
  residuals with respect to the sinusoidal modulation in units of
  $\sigma$. The values of the residuals corresponding to the first set
  of parameters are shown in the third column of Table \ref{Tab_opt}.
  The best-fit curve corresponding to the second set of parameters
  (hereafter best-fit 2) is shown in Fig. \ref{pino2} (top panel) with
  a solid line. In the bottom panel of Fig.  \ref{pino2}, we show the
  residuals with respect to the sinusoidal modulation in units of
  $\sigma$. The values of the residuals corresponding to the second
  set of parameters are shown in the fourth column of
  Table \ref{Tab_opt}.  We checked our results fitting the X-ray
  time-lags with respect to the best-fit parabola giving the X-ray
  ephemeris.  In this case, the residuals do not show any sinusoidal
  modulation, as expected. Fitting them with a constant, we found a
  $\chi^2(d.o.f.) = 42.8(29)$.


  The two best-fit sinusoidal functions indicate that the optical/UV
  eclipse lags the X-ray eclipse with a time shift of either  $161 \pm
  24$ s or $105 \pm 24$, which are significant at a confidence level
  of 6.7$\sigma$ and 4.4$\sigma$, respectively.  These values are
  compatible with the time-lag predicted by \cite{Hellier_mason1989}, 
 of $\sim 200$ s. Our most intriguing result,
  never previously  detected, 
is that the optical/UV eclipse times may oscillate
  in time with an amplitude of either  $194 \pm 29$ s (best-fit 1) or
  $267 \pm 43$ s (best-fit 2), these values being significant at a
  confidence level of 6.7$\sigma$ and 6.2$\sigma$, respectively.  The
  detected periods are $\sim 18$ yr (best-fit 1) and $\sim 283$ days
  (best-fit 2).  Both the detected periods are very long and 
  difficult to explain by invoking a superhump phenomenon \citep[see][and
  reference therein]{wang2010}.  The superhump excess $\epsilon$ is
  defined as $P_{sh}/P_{orb}-1$, where $P_{sh}$ is the superhump
  period, and also as $\epsilon = 0.18 q + 0.29 q^2$
  \citep{patterson2005}, where $q$ is the mass ratio $m_2/m_1$ with
  $m_1$ the neutron-star mass. Since the mass function and inclination
  angle of X1822-371 is well known \citep{Jonker_2001}, assuming a
  neutron-star mass of 1.4 M$_{\odot}$ and an inclination angle of
  87$^{\circ}$ we find that $q\simeq 0.29$ and $\epsilon \sim 0.077$.
  Consequently the superhump period should be $P_{sh}=1.077 P_{orb}
  =21598.38$ s.  A possible beat phenomenon between the superhump
  period and the orbital period could produce a period given by
  $(1/P_{orb}-1/P_{sh})^{-1} \simeq 3.25$ days.  This value is shorter
  than the two periodicities that we  reported. 

   We found that the lag changes with time and that this
    variation is compatible with a sinusoidal modulation at two
    different periods of 18 yr and 283 days, respectively. However, we
    cannot exclude shorter periodicities, but our data set of only  36
    optical/UV eclipse times spanning a time  of 10,000 days do not
    allow a rigorous study.  To investigate this aspect, long
    optical/UV observations of X1822-371 covering several contiguous
    orbital periods of the source would be necessary.

Finally we note that both these best-fit curves are strongly driven by
  the optical measures corresponding to orbital cycles 7,243 and 7,600
  and by the recent two UV eclipse arrival times obtained with HST and
  reported by BA10; these last two points are at orbital cycle 35,387
  and 35,395.

\section{Conclusions}

We have revisited and discussed the X-ray and optical/UV ephemerides of
X1822-371.   Fitting simultaneously the optical/UV and X-ray time
  delays, we have found that the optical/UV eclipses of X1822-371 lag
  the X-ray eclipses by $ 127 \pm 52$ s with a significance level of
  2.4 $\sigma$.  However, this time-lag may not be  constant in time.
   Fitting the optical/UV time-lags, we have found  a statistically
significant variation, which is compatible with a sinusoidal
  modulation at two different periods, $\sim
  18$ yr and 239 d. In the first case, the optical/UV eclipses lag the
  X-ray eclipse by an average time of $161 $ s (significance
  6.7$\sigma$) and this delay oscillates in time around this value
  with an amplitude of $194$ s (significance 6.7$\sigma$). In the
  second case, the optical/UV eclipses lag the X-ray eclipse by an
  average time of 105 s (significance 4.4$\sigma$), and this delay
  oscillates in time around this value with an amplitude of $267$ s
  (significance 6.2$\sigma$).

  Owing to  the relatively small number of points over a long-time span
  of 30 yr, we cannot be sure of the period of this modulation, because
  we cannot exclude much shorter periods. Long and relatively
  continuous optical/UV observations are necessary to prove or
  disprove the presence of this periodicity in the optical/UV
  eclipse time-lags. 

Our results confirm the value of the orbital solution derived
by the X-ray eclipse times given by BU10 and that the orbital period
derivative is three orders of magnitude larger than  expected
on the basis of the  conservative mass transfer driven by magnetic braking and
gravitational radiation. We have also confirmed this
result by combining the X-ray data and the optical/UV data of
X1822-371.

\begin{acknowledgements} 
  We thank the anonymous referee for the useful suggestions and
 A. Bayless for the fruitful interaction.
  authors acknowledge financial contribution from the agreement
  ASI-INAF I/009/10/0. AP acknowledges financial support from the
  Autonomous Region of Sardinia through a research grant under the
  program PO Sardegna ESF 2007-2013, L.R.7/2007,
  ”Promoting scientific research and technological innovation in
  Sardinia. LB and TD acknowledge support from the European Community's
  Seventh Framework Program (FP7/2007-2013) under grant agreement
  number ITN 215212 Black Hole Universe.
\end{acknowledgements} 
\bibliographystyle{aa} 
\bibliography{citations}

\begin{thebibliography}{12}
\expandafter\ifx\csname natexlab\endcsname\relax\def\natexlab#1{#1}\fi

\bibitem[{{Bayless} {et~al.}(2010){Bayless}, {Robinson}, {Hynes}, {Ashcraft},
  \& {Cornell}}]{Bayless2010}
{Bayless}, A.~J., {Robinson}, E.~L., {Hynes}, R.~I., {Ashcraft}, T.~A., \&
  {Cornell}, M.~E. 2010, \apj, 709, 251

\bibitem[{{Burderi} {et~al.}(2010){Burderi}, {di Salvo}, {Riggio}, {Papitto},
  {Iaria}, {D'A{\i}}, \& {Menna}}]{Burderi_2010}
{Burderi}, L., {di Salvo}, T., {Riggio}, A., {et~al.} 2010, \aap, 515, A44+

\bibitem[{{Hellier} \& {Mason}(1989)}]{Hellier_mason1989}
{Hellier}, C. \& {Mason}, K.~O. 1989, \mnras, 239, 715

\bibitem[{{Hellier} {et~al.}(1990){Hellier}, {Mason}, {Smale}, \&
  {Kilkenny}}]{Hellier90}
{Hellier}, C., {Mason}, K.~O., {Smale}, A.~P., \& {Kilkenny}, D. 1990, \mnras,
  244, 39P

\bibitem[{{Hellier} \& {Smale}(1994)}]{hellier_smale_94}
{Hellier}, C. \& {Smale}, A.~P. 1994, in American Institute of Physics
  Conference Series, Vol. 308, The Evolution of X-ray Binariese, ed. {S.~Holt
  \& C.~S.~Day}, 535--+

\bibitem[{{Ji} {et~al.}(2011){Ji}, {Schulz}, {Nowak}, \& {Canizares}}]{Ji2011}
{Ji}, L., {Schulz}, N.~S., {Nowak}, M.~A., \& {Canizares}, C.~R. 2011, \apj,
  729, 102

\bibitem[{{Jonker} \& {van der Klis}(2001)}]{Jonker_2001}
{Jonker}, P.~G. \& {van der Klis}, M. 2001, \apjl, 553, L43

\bibitem[{{Koyama} {et~al.}(2007){Koyama}, {Tsunemi}, {Dotani}, {Bautz},
  {Hayashida}, {Tsuru}, {Matsumoto}, {Ogawara}, {Ricker}, {Doty}, {Kissel},
  {Foster}, {Nakajima}, {Yamaguchi}, {Mori}, {Sakano}, {Hamaguchi},
  {Nishiuchi}, {Miyata}, {Torii}, {Namiki}, {Katsuda}, {Matsuura}, {Miyauchi},
  {Anabuki}, {Tawa}, {Ozaki}, {Murakami}, {Maeda}, {Ichikawa}, {Prigozhin},
  {Boughan}, {Lamarr}, {Miller}, {Burke}, {Gregory}, {Pillsbury}, {Bamba},
  {Hiraga}, {Senda}, {Katayama}, {Kitamoto}, {Tsujimoto}, {Kohmura}, {Tsuboi},
  \& {Awaki}}]{Koyama2007}
{Koyama}, K., {Tsunemi}, H., {Dotani}, T., {et~al.} 2007, \pasj, 59, 23

\bibitem[{{Parmar} {et~al.}(2000){Parmar}, {Oosterbroek}, {Del Sordo},
  {Segreto}, {Santangelo}, {Dal Fiume}, \& {Orlandini}}]{parmar2000}
{Parmar}, A.~N., {Oosterbroek}, T., {Del Sordo}, S., {et~al.} 2000, \aap, 356,
  175

\bibitem[{{Patterson} {et~al.}(2005){Patterson}, {Kemp}, {Harvey}, {Fried},
  {Rea}, {Monard}, {Cook}, {Skillman}, {Vanmunster}, {Bolt}, {Armstrong},
  {McCormick}, {Krajci}, {Jensen}, {Gunn}, {Butterworth}, {Foote}, {Bos},
  {Masi}, \& {Warhurst}}]{patterson2005}
{Patterson}, J., {Kemp}, J., {Harvey}, D.~A., {et~al.} 2005, \pasp, 117, 1204

\bibitem[{{Takahashi} {et~al.}(2007){Takahashi}, {Abe}, {Endo}, {Endo}, {Ezoe},
  {Fukazawa}, {Hamaya}, {Hirakuri}, {Hong}, {Horii}, {Inoue}, {Isobe}, {Itoh},
  {Iyomoto}, {Kamae}, {Kasama}, {Kataoka}, {Kato}, {Kawaharada}, {Kawano},
  {Kawashima}, {Kawasoe}, {Kishishita}, {Kitaguchi}, {Kobayashi}, {Kokubun},
  {Kotoku}, {Kouda}, {Kubota}, {Kuroda}, {Madejski}, {Makishima}, {Masukawa},
  {Matsumoto}, {Mitani}, {Miyawaki}, {Mizuno}, {Mori}, {Mori}, {Murashima},
  {Murakami}, {Nakazawa}, {Niko}, {Nomachi}, {Okada}, {Ohno}, {Oonuki}, {Ota},
  {Ozawa}, {Sato}, {Shinoda}, {Sugiho}, {Suzuki}, {Taguchi}, {Takahashi},
  {Takahashi}, {Takeda}, {Tamura}, {Tamura}, {Tanaka}, {Tanihata}, {Tashiro},
  {Terada}, {Tominaga}, {Uchiyama}, {Watanabe}, {Yamaoka}, {Yanagida}, \&
  {Yonetoku}}]{Takahashi2007}
{Takahashi}, T., {Abe}, K., {Endo}, M., {et~al.} 2007, \pasj, 59, 35

\bibitem[{{Wang} \& {Chakrabarty}(2010)}]{wang2010}
{Wang}, Z. \& {Chakrabarty}, D. 2010, \apj, 712, 653

\end{thebibliography}
\end{document}